# *Ordering Interrogative Questions for Effective Requirements Engineering: The W6H Pattern*

Mujahid Sultan
Treasury Board Secretariat
Government of Ontario
Toronto, Canada
Mujahid.Sultan@ontario.ca

Andriy Miranskyy
Department of Computer Science
Ryerson University
Toronto, Canada
avm@ryerson.ca

*Abstract*— **Requirements elicitation and requirements analysis are important practices of Requirements Engineering. Elicitation techniques, such as interviews and questionnaires, rely on formulating interrogative questions and asking these in a proper order to maximize the accuracy of the information being gathered. Information gathered during requirements elicitation then has to be interpreted, analyzed, and validated. Requirements analysis involves analyzing the problem and solutions spaces. In this paper, we describe a method to formulate interrogative questions for effective requirements elicitation based on the lexical and semantic principles of the English language interrogatives, and propose a pattern to organize stakeholder viewpoint concerns for better requirements analysis. This helps requirements engineer thoroughly describe problem and solutions spaces.**

**Most of the previous requirements elicitation studies included six out of the seven English language interrogatives 'what', 'where', 'when', 'who', 'why', and 'how' (denoted by W5H) and did not propose any order in the interrogatives. We show that extending the set of six interrogatives with 'which' (denoted by W6H) improves the generation and formulation of questions for requirements elicitation and facilitates better requirements analysis via arranging stakeholder views. We discuss the interdependencies among interrogatives (for requirements engineer to consider while eliciting the requirements) and suggest an order for the set of W6H interrogatives. The proposed W6H-based reusable pattern also aids requirements engineer in organizing viewpoint concerns of stakeholders, making this pattern an effective tool for requirements analysis.**

## I. Introduction

The textbook definition of engineering is the creation of cost-effective solutions to a problem through the application of scientific knowledge [1]. Requirements Engineering (RE), similarly, is the application of engineering principles to solve problems and optimize solutions. Patterns and re-usability are the basic principles of engineering. The success of a system is measured in terms of the degree to which it fulfills the purpose for which it was created, selected or acquired. Software is created, selected, or acquired to fulfill stakeholders' requirements, termed as problem space. Requirements elicitation from stakeholders, whether humans or systems, is challenging [2] and becomes more complex and subjective when the system users are human[1].

Ambiguity is omnipresent in the description, understanding and interpretation of human needs by others (e.g., by requirements engineers). Moreover, needs change rapidly after systems are developed, selected, or acquired. The source of this ambiguity is the communication between the stakeholders and the requirements elicitor. Both rely heavily on language and linguistics [3], [4]. Therefore, it is extremely important that both use proper and correct linguistics so that communication is clear, and the requirements lead to the correct definition of the problem.

The elicitation of requirements is the first step in the RE process [5], [6] and understanding actor goals can lead to better system requirements [7]. One of the most important goals of elicitation is to determine the problem, which the system needs to solve. The purposes of requirements definition are to elicit requirements from stakeholders, separate needs from wants, consider constraints, and propose requirements acceptable by system stakeholders. The requirements engineer then creates a link between the problem space and the solution space (see [8], [9] for literature review).

Elicitation techniques, such as interviews and questionnaires [10], [2] rely on interrogative questions. If some interrogative questions are missing or not asked in the correct order, information might be lost, leading to subpar requirements. These facts are recognized by academic community [4], and further corroborated by our industrial experience: information is generally missing for proper requirements analysis due to gaps in description of stakeholder needs. This missing information becomes crucial during requirements

[1] If the user is a system then the requirements are comparatively easy to gather, since the system "knows" what it needs, based on well-defined artifacts, such as interfaces and automated control machinery.

analysis, especially prioritization [11], [12], since the decisions made using incomplete information are often suboptimal. Therefore, our **research questions** are as follows:

**RQ1**: How can the generation of interrogative questions used by requirements elicitation techniques, such as interviewing, questionnaire, survey and brainstorming, be improved?

**RQ2**: How should these interrogative questions be ordered?

**RQ3**: How should interrogative questions for iterative development and prioritisation be formulated?

The rest of this paper is structured as follows. In Section II, we discuss the basic set of English language interrogatives, interdependencies among these and their relation to existing requirements elicitation techniques. Based on these findings, we propose a framework for generation of questions during requirements elicitation and analysis in Section III. We discuss application of the framework in Section IV. Finally, Section V presents the conclusions and future work.

## II. Seven Lexicalized Categories of English Language Interrogative Questions and Interdependencies among these

We discuss English language interrogatives used in requirements elicitation in Section II.A, followed by the interdependencies of these interrogatives in Section II.B.

### *A. Seven Lexicalised Categories of English Language Interrogatives*

English language has seven basic categories of interrogatives (also known as interrogative words): *which*, *what*, *where*, *when*, *who*, *why,* and *how* [13]. An interrogative word is a function word used to “generate” an interrogative sentence (question). For example, interrogative sentence ‘why did the chicken cross the road?’ is formed using interrogative *why*.

In most world languages *who*, *what*, *which*, and *where* are the four basic lexemes called the “major four” [14]. Cysouw [15] further describes the typology of interrogative categories:

- Major categories: person (*who*), thing (*what*), selection (*which*), place (*where*);
- Minor categories: manner (*how*), time (*when*);
- Incidental categories: reason (*why*).

The requirements elicitation frameworks W5H [16] and 6WHV, 6W2HV, and 6W2H2V [17], [18] (designed to gather requirements from viewpoints[2] of stakeholders) focus on the structured generation of interrogative sentences. The actual sentences depend on the project and context, but all of the questions are “driven” by the interrogatives. An alternative approach is to formulate all possible questions for each and every product and project out there. For example, Miller [19] formulates a list of over 2000 questions to elicit non-functional requirements. Unfortunately, one cannot guarantee completeness of this list.

The abovementioned frameworks [16]–[18] use six out of the seven English interrogative words: *what*, *where*, *when*, *who*, *why*, and *how* (hence the term W5H). This focus comes from the field of journalism [20], in which W5H interrogatives are used to describe an event by answering what happened, when it happened, who was involved, where it happened, why it happened, and how it happened.

However, requirements elicitation/problem definition from stakeholders’ viewpoints is not an event. Requirements engineer needs to create a link between the needs of stakeholders, the problem space, and the solution space. Viewpoints are widely used in defining the problem space in the requirements engineering domain [8], [16]. The interrogative *which*, omitted in the W5H set, is required to establish priorities and creates a link between the problem and the solutions spaces.

Therefore, to answer **RQ1 and RQ3** we need to extend the W5H set with *which* to make it W6H. It is tempting to substitute *which* with *what* or *who*, as often happens during informal conversations. However, in formal settings, doing so leads to the loss of information: *which* quantifies selection, whereas *what* is infinite. We provide examples in Section III.

### *B. The Order of the Interrogatives*

An additional issue arises because elicitation frameworks [16]–[18] do not consider the order and interdependencies among the interrogatives. In the English language, however, these interrogatives have a precedence relationship. For example, in order to answer the question ‘How secure is a feature?’ one must ask ‘Which feature?’

Ginzburg [21] explains the head-driven phrase structure grammar, which is widely accepted as a framework and is regarded the most explicit description of the syntax and semantics of English language interrogatives in any era of English syntax [22]. The importance of this work is also recognized in applied fields, such as computational linguistics, e.g., to develop clarification strategies for human-robot [23] or human-human [24] dialogues. Note that clarification issue is omnipresent in requirements elicitation and analysis processes, as requirements engineers often do not understand or misunderstand the data collected from stakeholders.

Cysouw elaborates on Ginzburg’s work by presenting morphological and lexical analysis of English language interrogatives [13], [15] and describes the precedence relationship among interrogatives, as shown in Figure 1. Note that the interrogative *which* (discussed in Section II.A), excluded from

[2] We use ANSI/IEEE Standard 1471-2000 definitions of stakeholder views and viewpoints [23]. A view is a representation of a whole system from the perspective of a related set of concerns. A viewpoint defines the perspective from which a view is taken. In other words, a viewpoint is where you are looking from - the vantage point or perspective that determines what you see; a view is what you see.

W5H-based approaches, plays an important role in the precedence relationship among other interrogatives.

We can use this precedence relationship graph to answer **RQ2**. To the best of our knowledge, no research has been conducted to establish the impact of the precedence relationship among the English language interrogatives on describing RE/requirements elicitation from stakeholders' viewpoints. We discuss the order of interrogatives specific to requirements elicitation and analysis in detail in Section III.

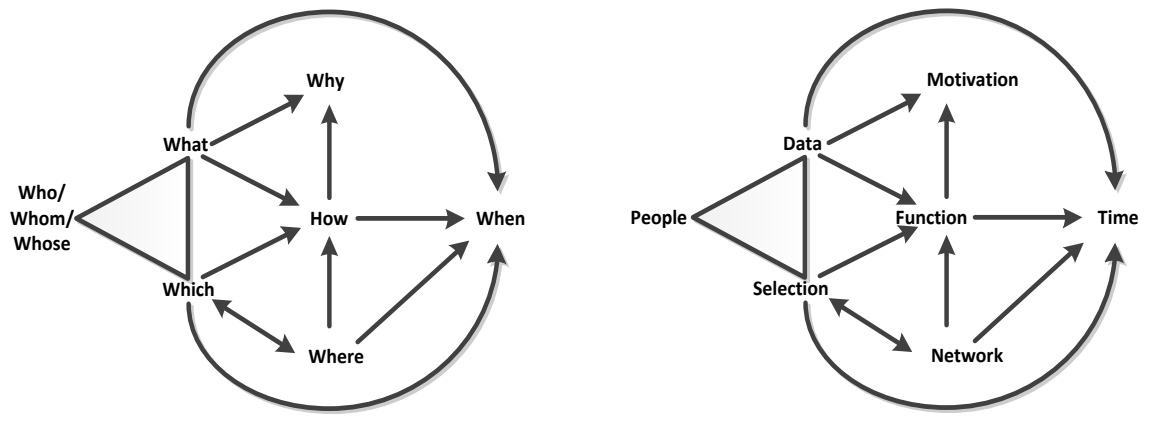

Fig.1. Left: Order and interdependencies of English interrogatives based on [13] and [22]. Right: material categories corresponding to the interrogative words. Legend: The edges represent relationships among interrogatives, and the arrows point to dependent interrogatives. Directionless edges indicate an absence of dependency among interrogatives. Bidirectional arrows indicate interrogatives have dependency upon each other.

## III. W6H PATTERN

Based on these findings, we introduce a pattern of interrogative questions for requirements elicitation (in the order given in first row of Table I), we denote this set of ordered interrogatives as W6H pattern and propose grouping by stakeholder viewpoint concerns to facilitate requirements analysis. The columns in Table I represent the ordered interrogatives. The rows in Table I are stakeholder groups (namely, Users, Developers, Legislators, and Decision-makers), drawing from the baseline requirements stakeholder taxonomy, proposed by Sharp et al. [25]. Each cell in Table I contains viewpoint concerns of a given stakeholder group which can be satisfied by asking the corresponding interrogative question. This pattern is inspired by use cases frequently encountered by the authors on multiple projects in large and complex enterprises.

In the following sub-sections, we give examples of how to reuse the W6H pattern, borrowing an example from each user group. We highlight the importance and usage of interrogatives' ordering and stress the usefulness of the interrogative *which* (missing from the W5H-based frameworks). Each subsection below elaborates on the following:

1) Use case of W6H pattern for a stakeholder in each group;
2) Importance of the order of interrogative questions;
3) Importance of *which* interrogative questions.

Note that the examples are neither exhaustive nor do they represent concerns of all users. For example, if the developers do not follow the Scrum methodology, the question 'In which Sprint should we deliver a given requirement?' would not be applicable to their processes. If they use an iterative process instead, they might ask the question 'In which iteration should we deliver a given requirement?'

### A. *Business Owner (member of Users group, responsible for software system justification)*

#### 1) *Use case of the Pattern*

Business owners are the most important stakeholders in the users' group. To capture the concerns of the business owners' we may pose the following interrogative questions, in the order given below, to minimize information loss and guarantee completeness:

- *Who* are the clients? For *whom* the system is being built? *Whose* needs are we fulfilling?
- In *what* business information (data) are these clients interested? *What* business information (data) will be used for *which* system?
- *Which* system will be deployed, and *where*? *Which* of the data will this system (iteration/release) utilize?
- *How* will the proposed system meet the needs of clients? (On *which* data it will operate and *how*?)
- *Why* was a specific business function/system provided at a specific location?
- *When* should this system be deployed and decommissioned? *When* should the system apply discounts to a set of customers?

#### 2) *Importance of the order of interrogative questions*

The order of questions is important in W6H pattern. For example, the question: 'Why our communication channels with clients are not efficient?' requires answering at least the following questions in the order given below:

- *Who* are our clients?
- *What* needs (data needs) do our clients have?
- *What* communication channels are we using?
- *Which* ones are not efficient?
- *Where* are our clients located?
- *How* are we reaching them?

Note that the questions are ordered according to Figure 1. We should not use the ordering given in this figure dogmatically, but it provides a convenient guide for determining the flow of questions to streamline gathering relevant data.

#### 3) *Importance of which interrogative questions*

What the system does, and why it does so, is captured in the business plan. A business plan describes the business's value proposition and customer segmentation. Typical value propositions include 'For whom we are creating the value?' and *'Which* ones of our customers' problems are we helping to solve?' Consider the following questions for service delivery channels to customers:

- *Which* customer needs are we satisfying?
- Through *which* channels are our customer segments reached?
- *Which* channels work the best?
- *Which* are the most cost-effective?

TABLE I. W6H-BASED PATTERN TO ORGANIZE STAKEHOLDER VIEWS AND INTERROGATIVE INVESTIGATIONS FOR EFFECTIVE REQUIREMENTS ELICITATION AND REQUIREMENTS ANALYSIS. ROWS ARE STAKEHOLDER GROUPS, COLUMNS ARE INTERROGATIVE QUESTIONS IN THE PROPOSED ORDER (WITH DEPENDENCIES, DISCUSSED IN FIGURE 1, SHOWN IN SUB-SCRIPT)) AND CELLS ARE HIGH LEVEL STAKEHOLDER CONCERNS FROM A VIEWPOINT.

| | **1**<br>**Who** | **2**<br>**What** | **3**<br>**Which** | **4**<br>**Where** | **5**<br>**How** $_{5 \rightarrow 2, 3/4}$ | **6**<br>**Why** $_{6 \rightarrow 2, 5}$ | **7**<br>**When** $_{7 \rightarrow 2, 3, 4, 5}$ |
|---|---|---|---|---|---|---|---|
| **Users Group** | -Primary users<br>-Secondary users<br>-Tertiary users<br>- Direct users<br>-Testers<br>-Business Continuity Planning (BCP) planners<br>-Disaster Recovery Planning (DRP) planners<br>-Continuity Of Operations Planning (COOP) planners | -Data elements<br>-Data sensitivity<br>-Data retention policy<br>-Data owners<br>-Data custodians<br>-Business rules<br>-Asset values<br>-Business function values<br>-Maximum Tolerable Down time (MTD) | -Which requirements are developed in current iteration/release<br>-Next iteration<br>-Which data is used by which business function<br>-MTD for business function<br>-Which system functions to be restored first in case of a disaster | -Business function locations<br>-Data locations<br>-Facility locations<br>-Recovery locations<br>-Re-construction locations | -Business function is carried<br>-Business scenario description<br>-User stories<br>-Functional Requirements (FRs)<br>-Non Functional Requirements (NFRs) | - Business need (which maps to business goals)<br>-Why a business function is provided at a specific location<br>-Why to have high value of a business functions (for DRP)<br>-Recovery Time Objective (RTO)<br>-Recovery Point Objective (RPO) | -Business cycles<br>-Business triggers<br>-Archiving polices<br>-Recovery times<br>-Reconstruc-tion times |
| **Developers Group** | -Requirements Engineers<br>-Architects<br>-Programmers<br>-Designers | -Entities and their relationships<br>-System functions and data<br>-Data entities<br>-Entity relationships<br>-Data retention<br>-Recovery planning<br>-Reconstruction planning | -Sprint<br>-Release<br>-Iteration<br>-Module<br>-Components<br>-Entities to FRs<br>-Entities to NFRs<br>-Entities to ACs/ADs<br>-CRUD matrix<br>-DRP selection<br>-BCP selection | -Network segments<br>-Where to deploy systems<br>-Where to deploy for DRP<br>-Where to reconstruct after disaster | -FRs<br>-NFRs<br>-Sub-systems<br>-Architectural Constraints (ACs)<br>-Architectural Descriptions (ASs)<br>-Patterns<br>-Security (CIA) concerns<br>-Privacy concerns | -Why to load balance a system<br>-Why replicated site, database<br>-Why to create demilitarized zone and extra fire walls<br>-Why RAID and which level<br>-Why to prepare for warm/cold recovery site | -Triggers<br>-Rules<br>-Scalability timelines (e.g. Xmas time)<br>-Availability timelines<br>-DRP consi-derations<br>-BCP consi-derations |
| **Legislators Group** | -Governance, Risk, and Compliance (GRC)<br>-Enterprise Architects<br>-Security<br>-Privacy<br>-Internal and external regulators<br>-Auditors (internal and external)<br>-Business Continuity & Disaster Recovery Planner<br>BC & DR planner | -Architectural patterns and standards<br>-Architectural-baseline<br>-Security standards and baseline<br>-Privacy standards<br>-Controls<br>-Performance indicators (KPIs) | -Architectural patterns and standards<br>-Architectural-baseline<br>-Security controls<br>-Privacy procedures<br>-Controls<br>-Key Performance indicators (KPIs)<br>-Controls for DRP/BCP<br>-Rules, regulations and standards | -The network controls (IDS, IPS)<br>-DRP sites<br>-COOP sites<br>- Recovery sites<br>-Reconstruction site | -NFRs<br>-Sub-systems<br>- ACs<br>-ADs<br>-Patterns<br>-Security (CIA)<br>-Privacy | -Sarbanes Oxley act (SOX) requirements<br>-NIST 800-34<br>-HIPPA requirements<br>-FIPPA requirements<br>-ISO 27000 series requirements<br>-Service Organization Controls (SOC) requirements | -Project governance timelines<br>-Enterprise governance timelines<br>-Audit timelines (internal and external) |
| **Decision-makers Group** | -Business managers<br>-Delivery managers<br>-CIOs<br>-CFOs<br>-CSOs | -Architectural patterns<br>-Architectural-baseline<br>-Security baseline<br>-Privacy standards<br>-Controls<br>-Performance indicators (KPIs) | - Performance indicators (KPIs)<br>- Controls for DRP/BCP<br>-Rules, regulations and standards | -Business locations<br>-System locations<br>-Data locations<br>-DRP site selection<br>-COOP site<br>-Reconstruction locations | -Key Performance reports<br>-Dashboards<br>-DRP testing | -Strategic goals<br>-Business objectives<br>-Opportunities and threats | - Project governance timelines<br>-Enterprise governance timelines<br>-Audit timelines (internal and external) |

To come up with the value proposition, key element of a business plan, we need to ask the seventh interrogative which, missing from the W5H-based approaches.

### *B. Requirements Engineer (member of Developers group tasked with bridging the gap between problem and solution space)*

#### *1) Use case of the Pattern*

Requirements engineer needs to understand and document the problem and solution space, Architectural Constraints (AC), and strategic goals of the organization. Following the W6H sequence of questions, given below, enables the requirements engineer to organize her concerns to look at the holistic picture:

- *Who* are the system users/actors? *Who* interacts with the system? *Whose* requirements are we fulfilling?
- *What* data elements will be required for the system? *What* relationships will data entities have? *What* (data) will be used by *which* application?
- *Which* application will be deployed *where*? *Which* functions are being built in this iteration/release? *Which* architectural components/patterns are being reused?
- *Where* (network segment) will an application (application component) be deployed? *Where* (data center location, network segment) will sensitive data be located?
- *How* will the application functions fulfill the requirements? *How* will the system meet the non-functional requirements? *How* will the quality metrics be met?
- *Why* does an application/database need to be replicated? *Why* are different network segments being created (e.g., some with more security than others)?
- *When* will an application apply discounts on *what* transactions (e.g., promotions based on business rules)? *When* will data need to be archived? *When* the capacity of a system needs to be scaled up/down?

#### *2) Importance of the order of interrogative questions*

To illustrate the importance of order in interrogatives, we take the example of access and authentication software system and evaluate the order in which the requirements engineer should gather information for requirements analysis. An access control system comprises of the following components and executed in the order given below.

1. Identification: *who* is to be authorized—person (*who*).
2. Authentication:
    a. Validation of credentials—data (*what*).
3. Authorization:
    a. This user has access rights to *which* systems.
    b. *Where* (network segment) these systems are located.
4. Access control:
    a. Method of and level of validation (for example single or two factor authentication) (*how*).
    b. *Why* a given user has certain level of access to the systems and *why* the user is authorized to have the access?
5. Accountability/Audit:
    a. *When* to monitor and log *what* actions of users (for audit purposes).

The importance of order is quite evident. One cannot ask *why* questions before answering *who*, *what*, *where* and *how* questions. Authorization cannot be established until identification and authentication are completed. Similarly, auditing and access levels cannot be controlled until identification, authentication, and authorization are completed.

#### *3) Importance of which interrogative questions*

A very important task of RE is to identify data required by system functions. Create, Read, Update and Delete (CRUD) matrices became prevalent with emergence of object oriented analysis and design [26]. CRUD matrices provide an easy mechanism to associate and link system functions with data elements. The '*which* interrogative question' is key in creating CRUD matrix. W5H-based approaches cannot create a CRUD matrix unless *which* interrogative question is introduced for the purposes of selection. In other words, we need to establish *which* data entities are to be used by *which* application functions. Although we might elicit requirements for entities and application functions separately, it is the '*which'* interrogative that establishes the link between data and functions.

In our experience, practitioners do use *which* interrogative questions to create the required link and to create CRUD matrices, but, due to lack of any formal methodology, it is not practiced consistently.

### *C. Business Continuity and Disaster Recovery Planner (member of Legislators group)*

#### *1) Use case of the Pattern*

Business Continuity Management (BCM) has two major parts, Business Continuity Planning (BCP) and Disaster Recovery Planning (DRP). BCP includes Business Impact Analysis (BIA) which is identification of a) critical business functions essential for the continuity of the business and b) key resources to the operations of these business functions (e.g., data, people, and locations). DRP involves a) Recovery strategy and b) Reconstruction Strategy. Based on ISO/IEC 27031 and ISO 22301 following are the steps of BCM:

1. Conduct BIA:
    - Select individuals for data gathering – *who*;
    - Identify critical business functions, and calculate maximum tolerable down time (MTD) – *what*;

- o Identify the resources these functions depend upon – *which*;
- o *Where* (the network segments/business locations) these functions and systems are located.

2. Identify preventive controls – *how*
3. Develop recovery strategies – *why*
4. Exercise drill – *when*
5. Maintain and update BCP - *when*

*2) Importance of the order of interrogative questions*

Note the order of the BCM steps above; to capture these BCP concerns the requirements engineer should pose interrogative questions in the order given in the Table I.

*3) Importance of which interrogative questions*

The first step in BCP/DRP involves the selection of business functions critical to the business (from among a finite set of all functions), demonstrating the need and necessity of the '*which* interrogative' for BCP/DRP. Typical questions to ask in BCP/DRP are: '*Which* systems are critical?' and '*Which* critical systems is it most important to recover first?' Answering these questions enables planning the disaster recovery site and type.

*D. CIO (member of Decision-makers group)*

*1) Use case of the Pattern*

The requirements engineer tasked to capture the concerns of the decision makers (e.g., the CIO of an organization) from the BCP and DRP perspectives should document the following:

- *Who* are the decisions makers? *Which* roles make decisions?
- *What* standards and controls are in place? *What* controls are the decisions makers interested in?
- *Which* controls and KPIs are they interested in?
- *Where* are the DRP sites? *Where* critical business functions are carried out in case of emergency? *Where* are the Continuity of Operations (COOP) sites located? *Where* will be the re-constructions site located?
- *How* are the Disaster Recovery Plans carried out (e.g., using full interruption test and/or simulations)?
- The reason a specific recovery strategy is more important than the other (alignment with strategic business goals) – *why.*
- *When* to conduct a DRP test (e.g., full interruption test or simulation test)?

*2) Importance of the order of interrogative questions*

To keep the natural flow for this use case, the order (as given in Table I) enables the requirements engineer to capture all concerns.

*3) Importance of Which interrogative questions*

Similarly, Decision-makers' concerns are '*Which* Key Performance Indicators (KPIs) are associated with *which* systems?' and '*Which* systems are used to achieve a given strategic goal?' Decision-makers are also interested in knowing, e.g., 'Which system to provision *where*?' and '*Which* to recover first after a disaster?'

## IV. DISCUSSION

### *A. Usage of W6H Pattern*

The W6H pattern (defined in Section III and shown in Table I) can be used to formulate interrogative questions to be asked by a member of a stakeholder group. For example, Users are asked interrogative questions to elicit requirements, whereas, the architect can be asked interrogative questions to analyze the requirements and bridge the gap between problem and solution spaces. Both requirements elicitation and gap bridging are widespread challenging tasks. The W6H pattern offers a structured approach to tackle these challenges, as we have shown in the use cases for every stakeholder group in Section III.

This pattern can be easily integrated into existing requirements elicitation processes and frameworks (see [2], [10] for literature review). As discussed in Section III, we provide frequently occurring viewpoint concerns in Table I. If a viewpoint concern, of the interest to a given requirements engineer, is not found in Table I, the requirements engineer can easily add it to the cells of Table I. In order to decide which cell the concern should be inserted to, the engineer has to answer two questions: 1) 'which interrogative word should be used to generate an interrogative sentence (question) needed to address the concern?' and 2) 'which stakeholder group is interested in the concern and is capable of answering the interrogative sentence (question)?' By answering these two questions, the requirements engineer will be able to determine intersection of a column and a row in the W6H pattern, and, consequently, the cell in which the concern should be inserted. Note that in some cases there may be more than one group of stakeholders interested in a given concern: e.g., business continuity planning is of interest to both Users and Legislator groups.

### *B. Interrogative word 'which' and prioritization*

As discussed in Section III.B.1, interrogative *which* can be used to prioritize requirements, e.g., by answering '*which* functions are being built in this iteration/release?' This is applicable to any iterative and incremental process, e.g., Scrum or V-model. As mentioned in Section II.B, requirements engineers often prioritize requirements under uncertainty in the presence of incomplete information [11], [12]. By construction, the W6H pattern helps to generate questions and enables requirements engineer to ask these questions in the proper order, reducing the amount of incomplete information, hence reduction of uncertainty during prioritization process.

### C. *Interrogative word 'why' and ordering*

*Why* interrogative is one of the most important interrogative words in RE. For example, by asking '*why* this requirement is important/useful/needed?' question (denoted by **WQ**) we can capture assumptions and rationale underlying the requirement and even decide if the requirement should be implemented at all. Human resources are precious – we do not want to spend significant efforts on gathering information about requirements, only to find out later that it is not needed.

It is natural to pose the question: why in Figure 1 and Table I the *why* interrogative relies on other interrogatives? Why we cannot ask the WQ question right away? It so happens that before asking the WQ question, requirements engineer should have a crude understanding of the requirements. At the very minimum, the engineer should ask '*who* are the stakeholders interested in a given requirement?' and '*what* is a high-level summary of the requirement?' The answer to the former is needed to identify the stakeholders who should respond to the WQ question; to the latter – to have a meaningful conversation with these stakeholders.

Upon gathering answers to these two questions, the engineer can finally ask the proper subset of stakeholders the WQ question. If, based on the answers to the WQ, the requirement is deemed important, then the engineer can invest her time and gather detailed information about the requirement, using the remaining questions from the W6H framework.

## V. Summary and Future Work

In this paper, we sought to answer the following research questions to improve information gathering during requirements elicitation and requirements analysis.

**RQ1**: How can the generation of interrogative questions used by requirements elicitation techniques, such as interviewing, questionnaire, survey and brainstorming, be improved?

**RQ2**: How should these interrogative questions be ordered?

**RQ3**: How should interrogative questions for iterative development and prioritisation be formulated?

By applying linguistic findings to answer **RQ1**, we demonstrated that extending the set of interrogatives of *what*, *where*, *when*, *who*, *why* and *how* (W5H) with *which* (W6H) improves the generation of questions for the requirements elicitation process. Answering **RQ2**, we showed that asking questions based on the W6H pattern (in the order of precedence described in Figure 1 and Table I) improves information flow for both requirements elicitation and requirements analysis. We also discussed that the '*which*' interrogative enables selection and prioritisation of requirements, answering **RQ3**.

Finally, we created Table I based on the W6H pattern for requirements analysis and for the formulation of questions to capture stakeholder viewpoint concerns during requirements elicitation. Table I serves as an effective tool for requirements analysis and bridges the gap between problem and solution spaces. The requirements engineer can use the pattern to analyse the problem and solutions spaces, assess any missing information, return to a shareholder's specific viewpoint concern and seek further information. We also described how additional stakeholders' concerns can be added to the pattern.

We presented use cases of the pattern for a member from every stakeholder group and demonstrated that the pattern guarantees the complete capture of stakeholders' viewpoint concerns, equipping the requirements engineer to perform effective requirements engineering. We showed that not following the pattern might leave gaps in stakeholder viewpoint concerns, leading to subpar requirements analysis and engineering. We also stressed that the missing *which* in W5H-based patterns plays an important role in the selection process and provides effective mechanisms for iterative, agile developments.

We believe that our findings are of interest to practitioners who can readily use the W6H pattern (either by itself or as part of an existing requirements elicitation and analysis process or framework) to generate questions for requirements elicitation and to improve the requirements analysis and prioritisation process. The findings will also be of interest to theoreticians. The linguistic theories provide a sound foundation for the extension and generalisation of our framework, enabling novel work in requirements elicitation and analysis.

These findings are based on the authors' two decades experience in numerous large-scale, complex enterprise projects in the public and private sectors. In this paper, we gave pedagogical examples of the W6H pattern usage based on real-world use cases frequently encountered by the authors. We plan to formally validate the pattern using datasets collected from industrial projects.

Stakeholder viewpoints and interrogative investigations based on W5H are frequently used in other domains, such as enterprise architecture (e.g., the Zachman Framework), which face similar issues in the missing *which* interrogative and the ordering of interrogatives. We plan to extend the W6H pattern to these domains.